\documentclass{elsart}
\usepackage{graphicx,amssymb}

\begin{document}

\begin{frontmatter}

\title{Some thoughts on theoretical physics}

\author{Constantino Tsallis}
\ead{tsallis@cbpf.br}

\address{Centro Brasileiro de Pesquisas Fisicas\\
Rua Xavier Sigaud 150, 22290-180 Rio de Janeiro -- RJ, Brazil 
}
\vspace{1cm}
\hspace{1.3cm}$\;\;\;\;\;\;\;\;\;\;\;\;\;\;\;\;\;\;\;\;\;\;\;\;\;\;\;\;\;\;\;\;$\small{{\it Si l'action n'a quelque splendeur de libert\'e, \\
\hspace{1.3cm}$\;\;\;\;\;\;\;\;\;\;\;\;\;\;\;\;\;\;\;\;\;\;\;\;\;\;\;\;\;\;\;$elle n'a point de gr\^ace ni d'honneur.}}\\
		\hspace{1.3cm}$\;\;\;\;\;\;\;\;\;\;\;\;\;\;\;\;\;\;\;\;\;\;\;\;\;\;\;\;\;\;\;\;\;\;\;\;\;\;\;\;\;\;\;\;\;\;$Michel de Montaigne [{\it Essais}, L. III, Chap. 9]

\date{December 28, 2003}

\begin{abstract}
Some thoughts are presented on the inter-relation between beauty and truth in science in general and theoretical physics in particular. Some conjectural procedures that can be used to create new ideas, concepts and results are illustrated in both Boltzmann-Gibbs and nonextensive statistical mechanics. The sociological components of scientific progress and its unavoidable and benefic controversies are, mainly through existing literary texts,  briefly addressed as well\footnote{Short essay based on the plenary talk given at the International Workshop on {\it Trends and Perspectives in Extensive and Non-Extensive Statistical Mechanics}, held in November 19-21, 2003, in Angra dos Reis, Brazil.}.  
\end{abstract}

\begin{keyword}
Semiotics \sep Boltzmann-Gibbs statistical mechanics \sep Nonextensive statistical mechanics \sep Entropy \sep Sociology of science

\end{keyword}

\end{frontmatter}

\section{Introduction}

In the context of semiotics --- the study of signs --- the American philosopher Charles Sanders Peirce [1839-1914] addressed three basic forms of inference, namely {\it abduction}, {\it induction} and {\it deduction}. We all use these intellectual operations to attribute {\it meanings} to what our senses perceive. We use them for connecting signs, hence for doing science. According to Pierce, {\it ``we think only in signs"} and {\it ``nothing is a sign unless it is interpreted as a sign"}. Let us illustrate the above  three concepts through some simple examples.

From 

{\it All stones in box $A$ are black.\\
Stone $S_j$ is black.}

we may infer

{\it Stone $S_j$ is from box $A$.}

The third line does not necessarily follow from the first two, it is but a conjecture. If, however, we assume it, we are doing {\it abduction}, Sherlock Holmes' favorite operation in his art of seeking for the ``relevant details", the art of creating plausible scenarios. 

From

{\it Stone $S_1$ is from box $A$ and it is black.\\
Stone $S_2$ is from box $A$ and it is black.\\
Stone $S_3$ is from box $A$ and it is black.\\
...}

we may infer

{\it All stones in box $A$ are black.}

This of course is {\it induction}, simple induction, not {\it complete induction}, i.e., Peano's sophisticated set of axioms, currently admitted to lead to a logically necessary consequence.

Finally, we have the familiar {\it deduction}, constantly used in the construction of theorems. For example, from

{\it All stones in box $A$ are black.\\
Stone $S_i$ is from box $A$.}

we may infer

{\it Stone $S_i$ is black}.

Although by no means always consciously, we use these three forms of inference to make progress in science, in all sciences, and in particular in theoretical physics. In fact, in one way or another, we use them to follow the ``royal road" for discovery, i.e., to make  {\it metaphors}.  

 ``The greatest thing by far is to be a master of metaphor. It is the
one thing that cannot be learned from others; it is also a sign of
genius, since a good metaphor implies an eye for resemblance." 
wrote Aristotle in his {\it Ars Poetica} [322 AC]. And many --- perhaps virtually all --- scientists, conscious or unconsciously, take as granted that without methaphors, no scientific progress would exist. One may go even further: Without metaphors {\it that have some form of beauty}, no efficient progress in science would exist, no new ideas would emerge! 
 
\section{Boltzmann-Gibbs statistical mechanics}
Let us illustrate, within the Boltzmann-Gibbs (BG) statistical mechanical frame, how the above concepts may act in inter-twingled manners. Let us start with probabilities. 

\subsection{Probabilities and percolation}
The operation AND corresponds to a {\it series} array, as indicated in Fig. 1(a). Information must flow through the first bond {\it and} also through the second bond, so that the two terminals are connected. Assuming that the bonds are independent, the composition law is given by
\begin{equation}
p_s=p_1p_2 \;\;\;\;(series)\;,
\end{equation}
where the subindex $s$ stands for {\it series}. 
\begin{figure}
\begin{center}
\includegraphics[width=9.0cm,angle=0]{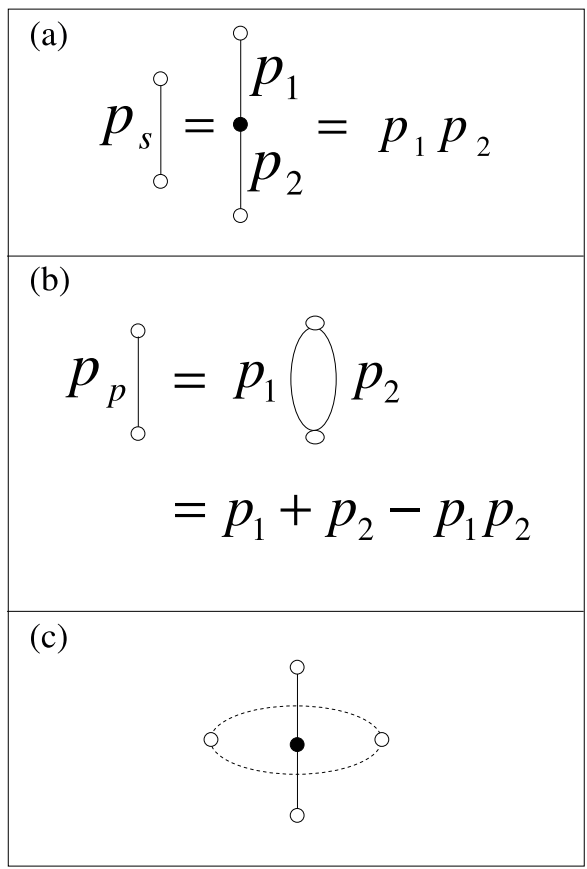}
\end{center}
\caption{\small Independent bonds. (a) Series array; (b) Parallel array; (c) Dual arrays, the dashed one cutting the solid-line one in such a way that every bond of any of those arrays is cutted by one and only one bond of the other array. See \cite{aglae} for details.}
\end{figure}
The operation OR corresponds to a {\it parallel} array, as indicated in Fig. 1(b). Information must flow through the first bond {\it or} through the second bond, so that the two terminals are connected. The composition law is given by
\begin{equation}
p_p=p_1p_2 +p_1(1-p_2)+p_2(1-p_1)=p_1+p_2-p_1p_2\;\;\;\;(parallel) \;,
\end{equation}
where the subindex $p$ stands for {\it parallel}. We may say that Eq. (2) is not beautiful enough, and can rewrite it as follows:
\begin{equation}
1-p_p=(1-p_1)(1-p_2) \;\;\;\;(parallel)\;.
\end{equation}
{\it Now} it may be considered as beautiful. Indeed, the parallel algorithm now appears identical to the series one (i.e., the product), and the functional forms associated with $p_p$, $p_1$ and $p_2$ are one and the same. This suggests the definition of a {\it dual} variable, namely
\begin{equation}
p^D \equiv1-p \;\;\;\;(duality)\;,
\end{equation}
where $D$ stands for {\it dual}; see Fig. 1(c). Now, Eq. (2) can be written in a beautiful form, namely
\begin{equation}
p^D_p=p^D_1p^D_2 \;\;\;\;(parallel)\;. 
\end{equation}
A basic question imposes itself: {\it By doing such operations have we obtained or can we obtain any scientific progress?}. The answer is {\it yes!}  We have created a scheme where abduction becomes almost a {\it must}. Let us illustrate this. Given the fact that the square lattice is a (topologically) self-dual array, it becomes kind of {\it natural} to expect for its bond-percolation threshold $p_c$ the following relation:
 \begin{equation}
p_c=p^D_c \;,
\end{equation}
hence $p_c=1-p_c$, hence $p_c=1/2$, which is well known to be the {\it exact} answer! (see references in \cite{aglae}).
\subsection{$Q$-state Potts model}
Let us consider now the one-bond Potts Hamiltonian
\begin{equation}
\mathbb{H}= -QJ \delta_{S_A,S_B} \;\;\;\;(S_A,S_B=1,2,...,Q) \;,
\end{equation}
where $Q>0$ is the number of states per spin\footnote{The usual notation is $q$, but we shall not use it here, in order to avoid confusion with the entropic index $q$ to be defined in a little while.}, and $J$ is a coupling constant ($J>0$ and $J<0$ respectively correspond to the ferromagnetic and antiferromagnetic interactions). The operation AND is as indicated in Fig 2(a). Seen from the outside world, and {\it if the system is in thermal equilibrium}, the flow of thermostatistical information in the series array of this figure corresponds to the following mathematical operation:
\begin{equation}
\sum_{S^\prime=1}^Q e^{-\beta Q (J_1 \,\delta_{S_A,S^\prime}+J_2\, \delta_{S_B,S^\prime})} \propto e^{-\beta Q \,J_s\, \delta_{S_A,S_B}}\;\;\;\;(\beta\equiv 1/kT) \;,
\end{equation}
where, as before, $s$ stands for {\it series}. 
This relation straightforwardly implies 
\begin{equation}
t_s=t_1 t_2 \;\;\;\;(series) \;,
\end{equation}
where
\begin{equation}
t \equiv \frac{1-e^{-Q \beta J}}{1+ (Q-1)e^{-Q \beta J}} \;\;\;\;(transmissivity)
\end{equation}
is referred to as the {\it thermal transmissivity} (or just {\it transmissivity}\footnote{More than two decades ago, during the short car drive from CBPF to my home in Rio de Janeiro, having Robin Stinchcombe as a guest, I asked him what sounded better in English, {\it transmittivity or transmissivity?} He answered that although both neologisms  sounded acceptable, {\it transmissivity} sounded ``more English". This is how this expression, since then frequently used in the specialized literature, was coined.}) of the bond. 
The limiting particular case $Q\to 1$ yields
\begin{equation}
t = 1-e^{-\beta J} \;,
\end{equation}
which precisely recovers the well known Fortuin-Kasteleyn isomorphism (see references in \cite{aglae}), thus transforming the Hamiltonian ferromagnetic problem into the bond percolation one.  The particular case $Q=2$ yields
\begin{equation}
t = \tanh (\beta J) \;
\end{equation}
which precisely is the high-temperature-expansion variable for the spin $1/2$ Ising model. 
\begin{figure}
\begin{center}
\includegraphics[width=8.5cm,angle=0]{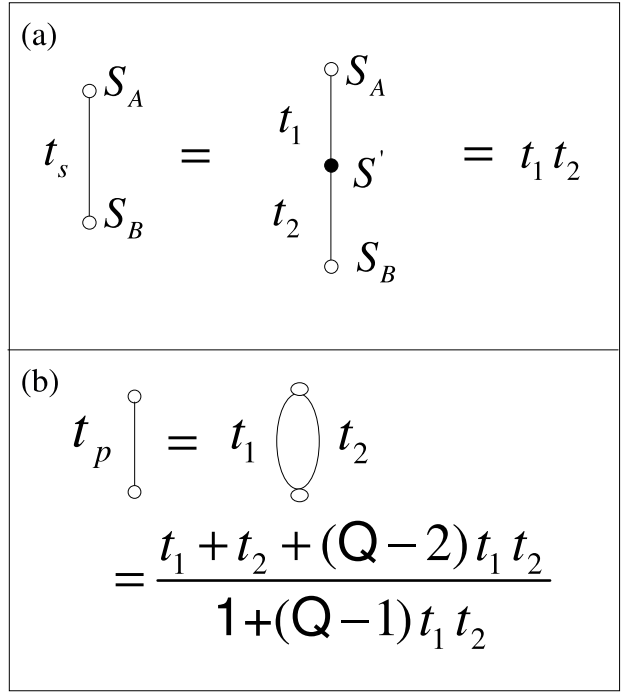}
\end{center}
\caption{\small $Q$-state Potts model. Each transmissivity $t$ is associated to a coupling constant $J$ through Eq. (10). (a) Series array; (b) Parallel array. See \cite{aglae} for details.}
\end{figure}

The operation OR corresponds to Fig. 2(b). This parallel algorithm is particular simple, namely 
\begin{equation}
J_p=J_1+J_2 \;\;\;\;(parallel) \;,
\end{equation}
or, equivalently,
\begin{equation}
t_p = \frac{t_1+t_2 +(Q-2)t_1t_2}{1+(Q-1)t_1t_2} \;\;\;\;(parallel) \;.
\end{equation}
Although this form is evidently more complex than Eq. (13), this is an interesting manner for writing the sum algorithm indicated in Eq. (13). Indeed, $t=1$ corresponds to full transmission of thermostatistical information ($\beta J \to \infty$), i.e., the two terminal spins are ``solidary" (or {\it collapsed}). We remark that, if we consider $t_1=1$ in Eq. (14), we obtain $t_p=1$ for {\it any} value of $t_2$. Interestingly enough, this is the same property that occurs in special relativity. More precisely, if we consider $Q=2$ and replace $(t_p,t_1,t_2) \to (v/c,v_1/c,v_2/c)$ ($c$ being the speed of light in vacuum), we obtain the well known composition of velocities in Einstein's special relativity. Although I have been unable to extract more from this mathematical feature, I consider it a beautiful metaphor, not to be dismissed!

Let us now go back along the lines we developed, in subsection 2.1, for probabilites. Eq. (14) can be re-written in a more beatiful form, namely,
\begin{equation}
\frac{1-t_p}{1+(Q-1)t_p} = \frac{1-t_1}{1+(Q-1)t_1}  \, \frac{1-t_2}{1+(Q-1)t_2} \;\;\;\;(parallel) \;, 
\end{equation}     
which suggests the following definition for dual variable:
\begin{equation}
t^D \equiv \frac{1-t}{1+(Q-1)t} \;\;\;\;(duality) \;.
\end{equation}
Eq. (15) can therefore be written in a very elegant form, namely
\begin{equation}
t^D_p=t^D_1 t^D_2 \;\;\;\;(parallel) \;.
\end{equation}
As a trivial application, one expects for the square-lattice Potts ferromagnet critical point, the following generalization of Eq. (6):
\begin{equation}
t_c=t^D_c \;,
\end{equation} 
hence $t_c=\frac{1-t_c}{1+(Q-1)t_c}$, hence 
\begin{equation}
t_c=\frac{1}{\sqrt{Q}+1} \;\;\;\;(\forall Q) \;,
\end{equation}
which indeed is the {\it exact} answer! (see references in \cite{aglae}).  

It is possible to go one more step in ``compactification" or ``economy". Indeed, the result in Eq. (19) depends on $Q$. Is it possible to make this dependence to disappear? Duality enables a positive answer. Let us define the variable
\begin{equation}
s \equiv \frac{\ln [1+(Q-1)t]}{\ln Q} \;.
\end{equation}  
It has the remarkable property that it transforms under duality in a $Q$-independent manner, namely
\begin{equation}
s^D(t) \equiv s(t^D) = 1-s(t) \;\;\;\;(\forall Q) \;.
\end{equation}
It immediately follows that the critical point in Eq. (19) can be rewritten as follows:
\begin{equation}
s_c=1/2 \;\;\;\; (\forall Q) \;.
\end{equation}
Eq. (21) in some sense transforms ``flow of information" into ``non flow of information", ``presence" into ``absence". In particular, it transforms $s=1$ into $s=0$, and reciprocally\footnote{It is, at least for me, an intriguing coincidence the fact that, as I shall mention in Section 3, the transformation $\ln [1+(Q-1)t]$ present in Eq. (20) (that I introduced close to 25 years ago) happens to {\it precisely} be the one that transforms (as shown in \cite{tsallis} in 1988, in a completely different context) the nonextensive entropy $S_q$ into the extensive Renyi entropy $S_q^R$, $q$ playing the role of $Q$. In total analogy with the case under discussion, the composition law of $S_q$ for probabilistically independent systems {\it depends} on $q$, whereas that of the Renyi entropy does {\it not}.}.  

\subsection{Break-collapse method}
This point is a good opportunity to remind that the so called {\it Break-collapse method} (see references in \cite{aglae}) for handling arbitrary graphs of Potts --- and even more complex --- systems is also entirely based on this ``presence-absence" game applied onto conveniently chosen variables. This method replaces lengthy and cumbersome tracing calculations by simple topological operations, and has thus enabled a considerable amount of efficient real-space renormalization-group calculation of critical frontiers and exponents, and even of entire equations of states. It is a very remarkable simplification, which in some sense replaces ``calculus" by ``geometry". The operational convenience (and even some degree of physical intuition) of such methods is widely known:  Illustrous predecessors are Dirac's bra-ket notation, and Feynman's diagrams.    
\begin{figure}
\begin{center}
\includegraphics[width=10.5cm,angle=-90]{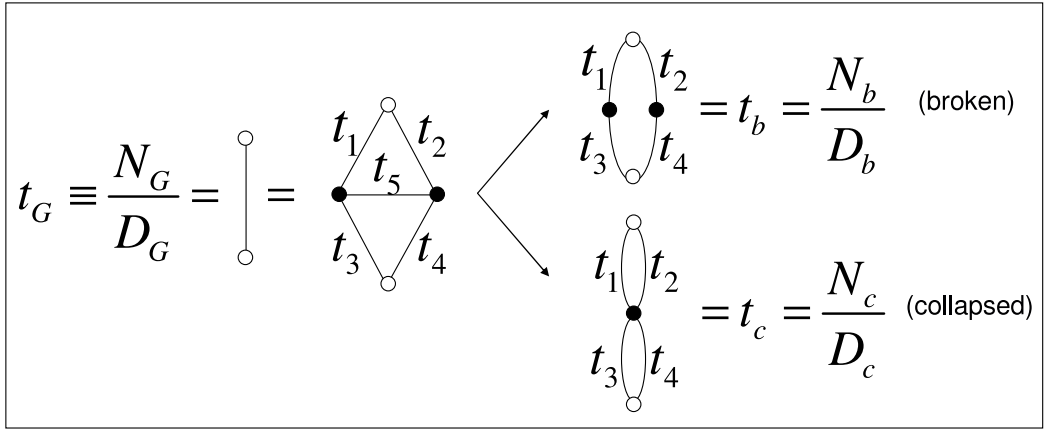}
\end{center}
\caption{\small Break-collapse operations on the Wheatstone-bridge Potts model. We start with the 5-bond graph, and arbitrarily choose bond $t_5$ to operate on.  We then obtain the two indicated 4-bond graphs by ``breaking" it (i.e., puting $t_5=0/1$) and ``collapsing" it (i.e., puting $t_5=1/1$). By applying the simple series and parallel composition laws for transmissivities we easily obtain $N_b$, $D_b$, $N_c$ and $D_c$. Replacing these into Eq. (23), we immediately have $N_G$ and $D_G$, the problem being thus solved {\it without explicitly performing any trace operations}. See \cite{aglae} for more details.}
\end{figure}

We shall illustrate now the procedure on the Potts magnet Wheatstone-bridge graph (which happens to be quite convenient for models defined on the isotropic ferromagnetic square lattice; see full details in \cite{aglae}). The basic break and collapse operations are shown in Fig. 3, where we have (arbitrarily) chosen to operate on bond-$5$. The composition algorithm is as follows:
\begin{eqnarray}
N_G=(1-t_5)N_b + t_5 N_c \nonumber \\
D_G=(1-t_5)D_b + t_5 D_c
\end{eqnarray}
 
The concepts of transmissivity, duality, break-collapse operations that I have briefly discussed in this Section have been devised on both rational and esthetical grounds. It seems legitimate to ask ``{\it  OK, this is a nice way to re-obtain results that are already known and which can, anyhow, be obtained through other, already existing, procedures. Is there more in it? Can we obtain new results? Can we make new predictions?}" 

The answer definitively is {\it Yes!}\footnote{New results, new predictions --- even sometimes {\it exact} predictions --- are only quite rarely obtained first on rigorous grounds. They are sometimes --- many times! --- just conjectures, but, as Galileo said, ``{\it Knowing with certainty a conclusion is not at all neglectable when one wants to discover the proof.}"}. For example, these procedures have enabled efficient theoretical discussions on the following systems, among others:

(i) Critical frontiers for the $Q$-state Potts ferro- and antiferro-magnets in bulk and surface, isotropic and anisotropic, square, triangular, honeycomb and more complex lattices;

(ii) Bond percolation threshold for the Kagome lattice: $p_c=0.52237207...$. The exact result is in fact still unknown, and it has been the subject of various studies by F.Y. Wu and collaborators, in addition to our own studies;

(iii) Random magnetism: Critical frontiers for dilute and mixed magnets;

(iv) Random resistor networks ($Q \to 0$);

(v) $Z(N)$ model;

(vi) Discrete and continuous $N$-vector model;

(vii) Polychromatic percolation;

(viii) Diffusion processes;

(ix) Fuzzy logic;

(x) Magnetism in quasi-crystals.

\subsection{On the Boltzmann-Gibbs entropy and thermal equilibrium distribution}
    
If $A$ and $B$ are two probabilistically independent systems, then
\begin{equation}
W(A+B)=W(A)W(B) \;,
\end{equation}
hence
\begin{equation}
\ln W(A+B)=\ln W(A)+ \ln W(B) \;. 
\end{equation}
Hypothesis (24) is practically satisfied at the stationary state (thermal equilibrium) of {\it isolated} systems whose microscopic dynamics is strongly chaotic (i.e., {\it positive}  Lyapunov exponents in their classical version). For such systems we expect equiprobability in Gibbs $\Gamma$-space to hold, hence the entropy
\begin{equation}
S_{BG}=k \ln W
\end{equation}
becomes perfectly adequate. It satisfies
\begin{equation}
S_{BG}(A+B)=S_{BG}(A)+S_{BG}(B) \;\;\;\;(extensiviy) \;.
\end{equation}
For arbitrary probabilities, Eq. (26) is generalized into
\begin{equation}
S_{BG}=-k \sum_{i=1}^W p_i \ln p_i \;.
\end{equation}
If 
\begin{equation}
p_{ij}(A+B)=p_i(A)\,p_j(B) \;\;\;\; (series) \;,
\end{equation} 
we can verify that Eq. (27), remarkably enough, {\it still remains true!}

If the (strongly chaotic) system is not isolated but in {\it thermal equilibrium} with a thermostat, then equiprobability in its $\Gamma$-space is violated, and it is replaced by the celebrated $BG$ distribution
\begin{equation}
p_i \propto e^{- \beta E_i} \;
\end{equation}
where $E_i$ is the $i^{th}$ eigenvalue of the  total energy of the system system. This distribution optimizes the entropy (28) under appropriate constraints. Equiprobability is recovered in the $\beta=0$ limit (i.e., $T \to \infty$), where in fact {\it all} statistics   (including the Fermi-Dirac and Bose-Einstein quantum ones) are expected to coincide. 

I mentioned here the well known expressions (28) and (30) with a specific purpose. It was (most probably) believed by Boltzmann and by Einstein \cite{einstein}, and it is nowadays believed by M. Baranger \cite{baranger}, E.G.D. Cohen \cite{cohen}, M. Gell-Mann (see Preface of \cite{gellmann}), myself, and surely many others, that such expressions should descend from dynamics (basically from $\vec{F}=m\vec{a}$, in the classical case), without further hypothesis than strong chaoticity. However, the rigorous steps that should provide Eqs. (28) and (30) starting from microscopic dynamics ... {\it are still unknown!} Nevertheless, no reasonable physicist would today contest the veracity of expressions (28) and (30), nor their predictive capabilities, just because we are still unable to deduce them rigorously. {\it How come it is so?, When did it become so?} Instead of addressing myself this interesting sociological phenomenon --- which is at the basis of all nontrivial progress in science, more specifically in all  substantial changes in basic scientific paradigms ---, I will quote the words of others.

William James wrote \cite{james}: ``Truth {\it happens} to an idea. It {\it becomes} true, is {\it made} true, by events."  

Spencer Weart wrote \cite{weart}: ``Sometimes, it takes a while to see what one is not prepared to look for."

Michael Riordan wrote \cite{riordan}: ``For [Charles Sanders] Pierce, the true hallmark of the ``real" is the observable consequences that a community of experienced practitioners agrees in actual practice." \\
And also \cite{riordan}: ``One of the great strengths of scientific practice is what can be called the ``withering skepticism" that is usually applied to theoretical ideas, especially in physics. We subject hypotheses to observational tests and reject those that fail. It is a complicated process, with many ambiguities that arise because theory is almost always used to interpret measurements. Philosophers of science say that measurements are ``theory laden", and they are. But good experimenters are irredeemable skeptics who thoroughly enjoy refuting the more speculative ideas of their theoretical colleagues. Through experience, they know how to exclude bias and make valid judgments that withstand the tests of time. Hypotheses that run this harrowing gauntlet and survive acquire a certain hardness --- or reality --- that mere fashions never achieve."

All this process took over 40 years (from say around 1870 to 1910-1915) before universal acceptance by the community of physicists that the revolutionary ideas (introduction of {\it probabilities} at the heart of physics) of Boltzmann were essentially correct. Reminding Max Planck's well known words \cite{planck} seems appropriate at this point:

``An important scientific innovation rarely makes its way by gradually winning over and converting its opponents: it rarely happens that Saul becomes Paul. What does happen is that its opponents gradually die out, and that the growing generation is familiarised with the ideas from the beginning." 

Or in Michael Fisher's version \cite{fisher}: ``However, I am afraid that in science, new and more correct ideas often win out only after their opponents die or retire. Evidently many people are not as open to rational conviction by new thoughts, as might be desirable!"

\section{Nonextensive statistical mechanics}

\subsection{The background}
First, a few chronological remarks, which I believe are typical of the emergence and evolution of scientific ideas. A Mexican-French-Brazilian workshop entitled ``First Workshop in Statistical Mechanics" was held in Mexico City, during 2 to 13 September 1985. I was acting as the coordinator of the Brazilian delegation; Edouard Brezin was acting as the coordinator of the French delegation. That was the time of fashionable multifractals and related matters. During one of the coffee breaks, everybody went out from the lecture room, excepting Brezin, a Mexican student (whose name I have not retained), and myself. I was just resting on my seat, and at some distance Brezin was explaining something to the student. At a certain moment, he addressed some point presumably related to multifractals --- from my seat I could not hear their conversation, but I could see the equations Brezin was writing. He was using $p^q$, and it suddenly came to my mind --- like a flash and without further intention --- that, with powers of probabilities, one could generalize standard statistical mechanics, by generalizing the $BG$ entropy itself and then following Gibbs' path. Back to Rio de Janeiro, I wrote on a single shot the expression for the generalized entropy, namely
\begin{equation}
S_q=k \frac{1-\sum_{i=1}^W p_i^q}{q-1} \;\;\;\;(S_1=S_{BG}). 
\end{equation} 
Although, at this early stage, there was no specific intention in this possible generalization, two things were clear: (i) on one hand, this possibility appeared as kind of natural to me since I had been thinking for years about the flow of information in graphs and systems within renormalization group schemes (see Section 2 of the present paper), and  (ii) on the other hand, the exponent $q$ would enable, as it does in multifractals, to focus on rather rare (or rather frequent) events that could be dominant in some physical phenomena. The case $q=1$ appeared then as the balanced, ``democratic" possibility. After writing the expression for the entropy, I studied a few of its properties, I found it kind of ``charming" (e.g., properties like positivity, concavity, equiprobability were all straightforwardly satisfied), and then I just stopped all that, occupied with various other projects.

Two years later, a workshop on cellular automata was held in Maceio-Brazil during 24 to 28 August 1987. The main organizer of the meeting, Enaldo F. Sarmento, left one afternoon free for any kind of discussions we would like to do in small groups. Then, for whatever reason, I just remembered the exotic entropy, and invited Evaldo Curado and Hans Herrmann to discuss with me about that. The discussion was done on the blackboard of an empty lecture room. Although none of us could really understand the possible physical relevance of that entropy, our exchange of ideas was quite lively, and both Evaldo and Hans were quite encouraging. So, I became once again stimulated by the idea. The day after, I took the plane back to Rio (before the end of the workshop in fact) and, during the flight, I scratched Gibbs' variational procedure on a sheet of paper. I obtained, for the stationary state distribution, the now quite known $q$-exponential form, namely\footnote{On the plane I obtained the form where $q-1$ plays the role of $1-q$. One form or the other is obtained for $p_i$, depending on how the energy constraint is written. It was only in Rio that I also found the other form, the one that was much later described in \cite{tsallis2}.}
\begin{equation}
p_i \propto [1-(1-q)\beta E_i]^{1/(1-q)} \;.
\end{equation}        
The idea was published in 1988 \cite{tsallis}. A few years later, Evaldo succeeded the connection to thermodynamics, which we published in 1991 \cite{tsallis2}. He arrived to the weighted form for the energy constraint, {\it independently} from my own calculation 3 or 4 years earlier. Later on, in 1998, Renio Mendes, Angel R. Plastino and myself published the form which is normally adopted nowadays \cite{tsallis3}. The first possible application in a physical system came from Angel Plastino Sr. and Angel Plastino Jr., who published in 1993 their by now well known paper on stellar polytropes \cite{plastinos}. During the     	
International Workshop on Nonlinear Phenomena, held in Florianopolis-Brazil during 7 to 9 December 1992, I had with me a copy of the preprint of the Plastino's paper. I discussed its content during hours with Roger Maynard. At the end of this long conversation, Roger and myself started being convinced that the whole game of this new entropy concerned the idea  of important nonlocal correlations between the elements of the system, beautifully illustrated through the long-range gravitational interaction. This same idea was reinforced by a --- also long --- conversation on the same subject that I had with Pierre-Gilles de Gennes in Catanzaro-Italy, during the International Conference on ``Scaling Concepts and Complex Fluids" held in 4 to 8 July 1994. 

The whole story of nonextensive statistical mechanics \cite{gellmann,review,review2} is quite long, and, during the last 15 years, crucial roles have been played by other protagonists, such as S. Abe, C. Anteneodo, M. Baranger, C. Beck, E.G.D. Cohen, V. Latora, M.L. Lyra, R.S. Mendes,  P. Quarati, A.K. Rajagopal, A. Rapisarda, A. Robledo, A.M.C. de Souza,     H.L. Swinney, F. Tamarit, G. Wilk, D.H. Zanette, and --- last but far from least --- M. Gell-Mann, whose inspiring input has had great influence.  The important contributions of all these, and others, might possibly be described in some other occasion. 

Now that some of the sociological context has been depicted --- even if very briefly --- , let me mention some important technical aspects. It can be easily shown that, if condition (29) ({\it series} configuration) is satified, then
\begin{equation}
\frac{S_q(A+B)}{k}=\frac{S_q(A)}{k} + \frac{S_q(B)}{k} + (1-q)  \frac{S_q(A)}{k}\frac{S_q(B)}{k}
\end{equation}
Notice that, if $q=1$, we can cancel $k$, but, if $q \ne 1$, we can {\it not}. This might well be related to the fact that $k$ is (together with $\hbar$, $c$ and $G$)  one of the four so-called universal constants of modern physics. In the same manner that $\hbar=1/c=G=0$ leads to the basic mechanics, namely that of Newton, the limit $1/k \to 0$ (or equivalently $(1-q)/k \to 0$) appears to also lead to some universal concept of loss or gain of physical information. This may be related to the fact that the Hawking entropy $S_{BH}$ per unit area $A$ for a black hole can be expressed exclusively in terms of these four constants [$S_{BH}/A=1/(4\hbar\, G\,k^{-1}\,c^{-3})$ ]. Notice also that, in the limit $(1-q)/k \to 0$, all statistics (those associated with Eq. (32), as well as the Fermi-Dirac and Bose-Einstein ones) exhibit confluence onto an universal form, namely that of the microcanonical ensemble, instance for which we have the minimal possible information on the system.

Along the lines explored in Section 2, we may expect that it should be possible to rewrite Eq. (33) in a $q$-independent form, by redefining the entropy. Indeed, this can be done as follows:
\begin{equation}
S_q^\prime(A+B)=S_q^\prime(A) + S_q^\prime(B)  \;\;\;\;(\forall q)\;,
\end{equation}
with
\begin{equation}
S_q^\prime \equiv k \frac{\ln [1+(1-q)S_q/k]}{1-q} \;.
\end{equation}
Interestingly enough, the Referees of the first version of my 1988 paper asked me to mention Renyi's entropy. It was already  in the manuscript! They had not noticed that...and neither had I! I was in fact unaware, at the time, of Renyi's entropy (and, in fact, of any other entropy whatsoever, different from the BG one): I just rediscovered  it independently in the form of $S_q^\prime$, by following the abductive arguments that I have used several times in the present essay. As already mentioned, it is interesting to notice the intriguing resemblance between Eq. (35) and Eq. (20).  		

At this stage, I should emphasize that one expects nonextensive statistical mechanics to play, for {\it weakly} chaotic systems (nonintegrable systems with {\it vanishing} Lyapunov exponents, in their classical version), a role similar to that played by BG statistical mechanics for {\it strongly} chaotic systems. Weakness of chaos appears to be consistent with important {\it nonlocal} space and/or time correlations. When such correlations are present in the system, it seems appealing that the concept of (lack of) information --- entropy --- be substantially modified. I then expect that vast classes of natural and artificial systems exist such that $S_q$ becomes {\it extensive} for a specific value of $q$, which adequately reflects the universality class of {\it nonlocality} ({\it absence of nonlocality}, i.e., {\it locality}, being the particular instance for which $q=1$, i.e., $S_{BG}$). Rephrasing, for systems with essential nonlocality, the entropy $S_q$ (with $q \ne 1$), which is {\it nonextensive} for independent systems, might (if $q$ is appropriately tuned onto that particular nonlocality) paradoxically become {\it extensive}! For instance, for equiprobability, $W \sim \mu^N\;(\mu>1)$ leads to $S_{BG} \propto N$; similarly, $W \sim N^\rho \; (\rho>0)$ leads to $S_{1-1/\rho} \propto N$. I like to imagine that such possibilities would definitively please Rudolf Julius Emmanuel Clausius, focused as he was on macroscopic thermodynamics!
 	
\subsection{About falsifiability and predictions}

Since the deep epistemological work by Karl Raimund Popper [1902-1994], it is clear that any scientific statement must be {\it falsifiable}. Let us illustrate this for nonextensive statistical mechanics. For a large class of mechanisms related to nonextensivity we expect anomalous diffusion such that
\begin{equation}
\langle x^2 \rangle \propto t^\gamma \;,
\end{equation}
$t$ being the time, and \cite{bukman}
\begin{equation}
\gamma =\frac{2}{3-q} \;.
\end{equation}
Stimulated by various colleagues, and very especially by a recent long conversation with Hans Herrmann in Salvador-Brazil, I present here a collection of systems whose corresponding results appear to be consistent with the above equation relating $\gamma$ and $q$.  

(i) {\it Hydra viridissima} \cite{arpita}

In experiments on cell aggregates of Hydra viridissima measuring the distribution of velocities, it was found
\begin{equation}
q_{stat} \simeq 1.50 \pm 0.05 \;,
\end{equation}
which, through Eq. (37), implies
\begin{equation}
\gamma \simeq 1.33 \pm 0.05 \;.
\end{equation}
Independent measurements of anomalous diffusion yielded
\begin{equation}
\gamma \simeq 1.23 \pm 0.1 \;,
\end{equation}
which is perfectly compatible with the prediction (39).

(ii) {\it Defect turbulence} \cite{daniels}

From the experimental velocity distribution it has been obtained
\begin{equation}
q_{stat} \simeq 1.50 \;,
\end{equation}
which, through Eq. (37), implies
\begin{equation}
\gamma \simeq 4/3 =1.33... \;.
\end{equation}
On the other hand, diffusion experiments have provided
\begin{eqnarray}
\gamma \simeq \mbox{1.16--1.50} \;,
\end{eqnarray}
which is compatible with prediction (42).

(iii) {\it Long-range Hamiltonian (HMF)}

Computational simulations yielded \cite{latora}
\begin{equation}
\gamma \simeq 1.38\;,
\end{equation}
hence one expects
\begin{equation}
q \simeq 1.55 \;.
\end{equation}
Indeed, from relaxation of the velocity correlation function, recent results \cite{anteneodotamarit,pluchino} provide
\begin{equation}
q_{rel} \simeq 1.50 \pm 0.15 \;,
\end{equation}
which is compatible with prediction (45).

(iv) {\it Finance (NYSE and NASDAQ)} \cite{osorio}

From stock return distributions it has been obtained
\begin{eqnarray}
q_{stat} \simeq     \mbox{1.38--1.43}\;, 
\end{eqnarray}
which implies
\begin{eqnarray}
\gamma \simeq \mbox{1.23--1.27} \;.
\end{eqnarray}
This (falsifiable) prediction remains to be checked. See \cite{tsalliscagliari} for details concerning the various values for the entropic index $q$ ($q_{stat}$, $q_{rel}$ and $q_{sen}$). 

In all these examples, {\it why have we only checked the validity of Eq. (37), and not independently predicted $q$ and $\gamma$?} The reason is quite elementary: Unless detailed studies reveal the microscopic (or at least mesoscopic) dynamics of such systems, it is not possible to calculate {\it a priori} the index $q$ (nor the exponent $\gamma$). Indeed, it has been profusely proved in the literature (\cite{dynamics} and references therein) that $q$ is determined by the universality class corresponding to the dynamics of the system. However, even when this dynamics happens to be unknown (or hardly tractable), we can still make consistency checks of the theory. Relation (37) precisely is one of such possible checks. 

\section{Some sociological aspects}

All changes of paradigm in science are sociologically complex, they follow paths that are by no means free from controversies, ambiguities, misunderstandings, reformulations and similar features --- even passion ---. Theoretical physics is no exception, very especially in what concerns those formalisms that use in an essential manner theory of probabilities, such as quantum mechanics and statistical mechanics. G. Nicolis and D. Daems wrote \cite{nicolis}: 

``It is the strange privilege of statistical mechanics to stimulate and nourish passionate discussions related to its foundations [...]."

Andrea Rapisarda reminded us the words of Arthur Schopenauer [1788 - 1860]:

`` All truths pass through three stages: First, they are considered ridiculous, second, they are violently adversed, third, they are accepted and considered self-evident."

Indeed, whether it represents or not a change in some relevant physical paradigm --- time will tell us! ---, nonextensive statistical mechanics seems to perfectly spouse the role. 

I would like to mention in this Section some of the most frequent or relevant points. For example, after my talk in the Vth Latin American Workshop on Nonlinear Phenomena and XIth Medyfinol Meeting held in Canela-Brazil during 28 september to 3 October 1997, the session chairwoman Marcia Barbosa provocatively exclaimed ``{\it I want a theory!}" This is an interesting and recurrent point. But, what is a theory? What is a {\it physical} theory? What is a formalism? What is a description? And what an explanation? Each one of these words --- that we scientists use all the time! --- is full of semantics, ambiguities and subjectivism.  I shall make no epistemological attempt of defining here these concepts. I will, however, reproduce Cassirer's words \cite{cassirer} [free translation]:\\
``When we ask to Mach and to Planck, to Boltzmann or to Ostwald, to Poincare or to Duhem, what a physical theory is, what is it that it can do, we obtain from them, not only answers that are different, but that are contradictory as well."\\
So, I will simply mention here what I understand as a {\it physical theory} as applied to nonextensive statistical mechanics. As for BG statistical mechanics, it would be extremely pleasant if we could derive, from dynamical first principles (e.,g., from Newtonian mechanics, or from quantum mechanics), the microscopic expression of the entropy in terms of probabilities, as well as the variational principle associated with that entropy. However, since this program has been fulfilled not even for the standard BG theory, there is little hope at the present moment that it becomes mathematically accessible for arbitrary value of $q$. In contrast, there is presently good hope that, for a vast class of thermostatistical systems, we will be able to deduce {\it a priori} the specific value of $q$ characterizing the universality class to which the system belongs. In fact, this fascinating program was only starting at the time of the Canela meeting, but it has considerably advanced since then, and nowadays we do know the answer for more than a dozen of classes of nonlinear dynamical systems. At this point, we may say that the $q \ne 1$ construct seems to be (quickly) approaching an epistemological status sensibly similar to that of the BG construct, whether one chooses to call it physical theory, formalism, description, explanation, or whatever. 

Another point which deserves mention at the sociological level is that several public debates have already occurred\footnote{The full list of formally organized public debates between only two opponents is, as far as I can tell, the following:

(i) Itamar Procaccia versus myself, in the III Gordon Research Conference on {\it Modern Developments in Thermodynamics}, held in Il Ciocco-Barga-Italy, during 18 to 23 April 1999.  \\
(ii) Roberto Luzzi gave, at my invitation, a critical long seminar at the Centro Brasileiro de Pesquisas Fisicas, Rio de Janeiro, in October 2002. It was immediately followed by a short reply by myself. As a consequence, a formal debate between Luzzi and myself was organized by the SBF (Brazilian Physical Society) in the presence of the 1200 participants at the XXVI Encontro Nacional de F\'\i sica da Mat\'eria Condensada, held in Caxambu, MG-Brazil, during 6 to 11 May 2003. Luzzi declined the formal invitation of the President of SBF. \\ 
(iii) Joel L. Lebowitz versus myself, in the Los Alamos National Laboratory International Workshop  on {\it Anomalous Distributions, Nonlinear Dynamics and Nonextensivity}, held in Santa Fe, NM-USA, during  6 to 9 November 2002. \\
(iv) Dieter H.E. Gross versus myself, in the American Physical Society March Meeting, held in Austin,TX-USA, during  3 to 7 March 2003.}. They undoubtedly make the scientific discussion and general understanding to become deeper and more precise.

Also, various controversial papers have been published either in peer-reviewed scientific journals or in automatic archives (e.g., the LANL ones). Specific replies to virtually all of them can be found in \cite{review2} (and in references therein).

As an opportune epilog of this Section, I would like to quote the old proverb:

``Do not fear the scientific fact, but the version of the political fact."

\section{Conclusion}

I would end with two literary pieces. Better than anything else, they express the main message I would like to convey in this talk.

From Jos\'e Saramago \cite{saramago}, fragments of {\it O conto da ilha desconhecida} (1997):
                                                 
o rei, com o pior dos modos, perguntou, Que \'e  que queres, \\
D\'a-me um barco, disse. \\
E tu para que queres um barco, pode-se saber, \\
Para ir \`a procura da ilha desconhecida, respondeu o homem, \\
Que ilha desconhecida, perguntou o rei disfar\c{c}ando o riso, \\
A ilha desconhecida, repetiu o homem, \\
Disparate, j\'a n\~ao h\'a ilhas desconhecidas,\\ 
Quem foi que te disse, rei, que j\'a n\~ao h\'a ilhas desconhecidas,\\ 
Est\~ao todas nos mapas, \\
Nos mapas s\'o est\~ao as ilhas conhecidas,\\ 
E que ilha desconhecida \'e essa de que queres ir \`a procura,\\ 
Se eu to pudesse dizer, ent\~ao n\~ao seria desconhecida, \\
A quem ouviste tu falar dela, perguntou o rei, agora mais s\'erio,\\ 
A ningu\'em, \\
Nesse caso, por que teimas em dizer que ela existe,\\ 
Simplesmente porque \'e imposs\'\i vel que n\~ao exista uma ilha desconhecida,\\ 
E vieste aqui para me pedires um barco, \\
Sim, vim aqui para pedir-te um barco, \\
E tu quem \'es, para que eu to d\^e, \\
E tu quem \'es, para que n\~ao mo d\^es,\\ 
Sou o rei deste reino, e os barcos do reino pertencem-me todos,\\ 
Mais lhes pertencer\'as tu a eles do que eles a ti, \\
Que queres dizer, perguntou o rei, inquieto, \\
Que tu, sem eles, \'es nada, e que eles, sem ti, poder\~ao sempre navegar, \\
E essa ilha desconhecida, se a encontrares, ser\'a para mim, \\
A ti, rei, s\'o te interessam as ilhas conhecidas, \\
Tamb\'em me interessam as desconhecidas quando deixam de o ser,\\ 
Talvez esta n\~ao se deixe conhecer, \\
Ent\~ao n\~ao te dou o barco, \\
Dar\'as.\\
................................\\
mas quero encontrar a ilha desconhecida, quero saber quem sou  eu quando nela estiver,\\
N\~ao o sabes, \\
Se n\~ao sais de ti, n\~ao chegas a saber quem \'es, Que \'e necess\'ario sair da ilha para ver a ilha, que n\~ao nos vemos se n\~ao nos sa\'\i mos de n\'os,\\
A ilha desconhecida \'e coisa que n\~ao existe, n\~ao passa duma id\'eia da tua cabe\c{c}a, os ge\'ografos do rei foram ver nos mapas e declararam que ilhas por conhecer \'e coisa que se acabou desde h\'a muito tempo,\\
Dev\'\i eis ter ficado na cidade, em lugar de vir atrapalhar-me a navega\c{c}\~ao, \footnote{{\bf Free translation}: {\it The tale of the unknown island.}

the king, in the worst manner, asked, What do you want,

Give me a boat, he said.

And what do you want a boat for, may we know,

To go to the search for the unknown island, the man answered,

What unknown island, asked the king hiding his lough,

The unknown island, the man repeated,

Nonsense, there no more unknown islands,

Who told you, king, that there are no more unknown islands, 

They are all in the maps,

In the maps there are only the known islands,

And what unknown island is that you want to search for,

If I could tell you, then it would not be unknown,

Whom from you heard about it, asked the king, now more serious,

From nobody,

In that case, why you insist in saying that it exists,

Simply because it is impossible that it does not exist an unknown island,

And you came here to ask me for a boat, 

Yes, I came here to ask you for a boat,

And who are you, for me to give it to you,

And who are you, for not give it to me,

I am the king of this kingdom, and the boats of the kingdom belong to me all,

More would you belong to them than they to you,

What do you mean, the king asked, uncomfortable,

That you, without them, are nothing, while them, without you, will always be able to navigate,

And that unknown island, if you find it, will be for me,

You, king, are only interested in the known islands,

I am also interested in the unknown ones when they cease being so, 

Maybe this one will not allow to be known,

Then, I will not give you the boat,

Yes, you will.

...................................

but I want to find the unknown island, I want to know who am I when I shall be there,

Don' t you know it,

If you do not leave out from yourself, you never arrive to know whom you are, That it is necessary to leave out from the island to see the island, that we do not see ourselves if we do not leave out from ourselves,

The unknown island is something that does not exist, it is not more than an idea in your head, the geographers of the king went to see in the maps and declared that islands yet to be known is something that finished long time ago,  

You should have stayed in town, instead of coming here to disturb my navigation,}\\

From Marco Bersanelli \cite{bersanelli}, {\it Sofia e la scoperta delle fragole} (1997):

A Gutenberg, tra le verdissime colline austriache, una mattina saliamo per il sentiero che attraversa il bosco scuro e profumato alle spalle del paese. Dopo mezz'ora di cammino troviamo sulla destra una sorgente presso una radura e ci fermiamo a bere. Con una grande espressione di felicit\`a ad un tratto Sofia, la piccola di tre anni, esclama: ``Mamma, mamma!! una fragola!!". Gli altri due accorrono e, constatato che la sorellina ha prontamente raccolto e inghiottito il frutto della sua scoperta, si mettono a cercare, presto seguiti dai genitori. ``Un'altra!" e dopo un po': ``Guarda qui, ce ne sono altre tre, quattro...". La caccia \`e aperta. Cercando in quel prato abbiamo presto riempito un bicchiere di fragole di bosco. Poi al ritorno, con mia sincera sorpresa, ripercorrendo lo stesso sentiero dalla sorgente in gi\`u ne abbiamo trovate altrettante! Zero fragole all'andata, forse un centinaio al ritorno: un effetto statisticamente schiacciante. Cos'era cambiato?\\
Eravamo cambiati noi. \footnote{{\bf Free translation}: {\it Sophia and the discovery of the strawberries.}

In Gutenberg, between the very green austrian hills, a morning we climb along the path that crosses the obscure and perfumed wood at the back of the countryside. After half an hour walk we found on the right side a fountain near a clear land and we stopped to drink. With a great expression of happiness suddenly Sophia, the three year old small one, exclaimed: ``Mammy, mammy!! a strawberry!!". The other two run and, having verified that their little sister had quickly taken and swallowed the fruit of her discovery, start searching, quickly followed by the parents. ``Another one!", and after a while: ``Look here, there are three, four...". The hunting is open. By searching in that field we quickly filled a glass of wood strawberries. On the way back, to my sincere surprise, following once again the same path down from the fountain we found many more! Zero strawberries along our way up, perhaps one hundred along our way down: an effect statistically enormous. What had changed?\\
We did.}

\section*{Acknowledgments}
The present talk is dedicated to my parents Emmanuel Tsallis and Cleopatra Yavassoglou, to my family, and to Plato [427-347 BC], who first discussed the connection between truth and beauty. \\
I express here my deep gratitude to the chairpersons Evaldo M.F. Curado, Hans J. Herrmann and Marcia Barbosa, as well as to all the other organizers who made possible  the wonderful meeting in Angra dos Reis during which this talk was presented. 
Partial support by FAPERJ, CNPq, PRONEX and FINEP (Brazilian agencies) is also gratefully acknowledged.

\end{document}